\documentclass[twocolumn,showpacs,prc,superscriptaddress]{revtex4}          
\def\snn{$\sqrt{s_{\rm NN}}$}

\newcommand{ \be }{\begin{equation}}       
\newcommand{ \ee }{\end{equation}}       
\newcommand{ \bea }{\begin{eqnarray}}       
\newcommand{ \eea }{\end{eqnarray}}

\newcommand{ \mean }[1]{\left\langle #1 \right\rangle}   
   
\usepackage{color}

\usepackage{graphicx} 
\usepackage{fixmath} 
\usepackage{dcolumn} 
\usepackage{bm} 
\usepackage{multirow}
 
\begin{document}          
\title{       
Fluctuations of charge separation perpendicular to the event plane and local
parity violation in \snn$=200$ GeV Au+Au collisions at RHIC
} 
 
\affiliation{AGH University of Science and Technology, Cracow, Poland}
\affiliation{Argonne National Laboratory, Argonne, Illinois 60439, USA}
\affiliation{University of Birmingham, Birmingham, United Kingdom}
\affiliation{Brookhaven National Laboratory, Upton, New York 11973, USA}
\affiliation{University of California, Berkeley, California 94720, USA}
\affiliation{University of California, Davis, California 95616, USA}
\affiliation{University of California, Los Angeles, California 90095, USA}
\affiliation{Universidade Estadual de Campinas, Sao Paulo, Brazil}
\affiliation{Central China Normal University (HZNU), Wuhan 430079, China}
\affiliation{University of Illinois at Chicago, Chicago, Illinois 60607, USA}
\affiliation{Cracow University of Technology, Cracow, Poland}
\affiliation{Creighton University, Omaha, Nebraska 68178, USA}
\affiliation{Czech Technical University in Prague, FNSPE, Prague, 115 19, Czech Republic}
\affiliation{Nuclear Physics Institute AS CR, 250 68 \v{R}e\v{z}/Prague, Czech Republic}
\affiliation{University of Frankfurt, Frankfurt, Germany}
\affiliation{Institute of Physics, Bhubaneswar 751005, India}
\affiliation{Indian Institute of Technology, Mumbai, India}
\affiliation{Indiana University, Bloomington, Indiana 47408, USA}
\affiliation{Alikhanov Institute for Theoretical and Experimental Physics, Moscow, Russia}
\affiliation{University of Jammu, Jammu 180001, India}
\affiliation{Joint Institute for Nuclear Research, Dubna, 141 980, Russia}
\affiliation{Kent State University, Kent, Ohio 44242, USA}
\affiliation{University of Kentucky, Lexington, Kentucky, 40506-0055, USA}
\affiliation{Institute of Modern Physics, Lanzhou, China}
\affiliation{Lawrence Berkeley National Laboratory, Berkeley, California 94720, USA}
\affiliation{Massachusetts Institute of Technology, Cambridge, MA 02139-4307, USA}
\affiliation{Max-Planck-Institut f\"ur Physik, Munich, Germany}
\affiliation{Michigan State University, East Lansing, Michigan 48824, USA}
\affiliation{Moscow Engineering Physics Institute, Moscow Russia}
\affiliation{National Institute of Science Education and Research, Bhubaneswar 751005, India}
\affiliation{Ohio State University, Columbus, Ohio 43210, USA}
\affiliation{Old Dominion University, Norfolk, VA, 23529, USA}
\affiliation{Institute of Nuclear Physics PAN, Cracow, Poland}
\affiliation{Panjab University, Chandigarh 160014, India}
\affiliation{Pennsylvania State University, University Park, Pennsylvania 16802, USA}
\affiliation{Institute of High Energy Physics, Protvino, Russia}
\affiliation{Purdue University, West Lafayette, Indiana 47907, USA}
\affiliation{Pusan National University, Pusan, Republic of Korea}
\affiliation{University of Rajasthan, Jaipur 302004, India}
\affiliation{Rice University, Houston, Texas 77251, USA}
\affiliation{Universidade de Sao Paulo, Sao Paulo, Brazil}
\affiliation{University of Science \& Technology of China, Hefei 230026, China}
\affiliation{Shandong University, Jinan, Shandong 250100, China}
\affiliation{Shanghai Institute of Applied Physics, Shanghai 201800, China}
\affiliation{SUBATECH, Nantes, France}
\affiliation{Temple University, Philadelphia, Pennsylvania, 19122, USA}
\affiliation{Texas A\&M University, College Station, Texas 77843, USA}
\affiliation{University of Texas, Austin, Texas 78712, USA}
\affiliation{University of Houston, Houston, TX, 77204, USA}
\affiliation{Tsinghua University, Beijing 100084, China}
\affiliation{United States Naval Academy, Annapolis, MD 21402, USA}
\affiliation{Valparaiso University, Valparaiso, Indiana 46383, USA}
\affiliation{Variable Energy Cyclotron Centre, Kolkata 700064, India}
\affiliation{Warsaw University of Technology, Warsaw, Poland}
\affiliation{University of Washington, Seattle, Washington 98195, USA}
\affiliation{Wayne State University, Detroit, Michigan 48201, USA}
\affiliation{Yale University, New Haven, Connecticut 06520, USA}
\affiliation{University of Zagreb, Zagreb, HR-10002, Croatia}

\author{L.~Adamczyk}\affiliation{AGH University of Science and Technology, Cracow, Poland}
\author{J.~K.~Adkins}\affiliation{University of Kentucky, Lexington, Kentucky, 40506-0055, USA}
\author{G.~Agakishiev}\affiliation{Joint Institute for Nuclear Research, Dubna, 141 980, Russia}
\author{M.~M.~Aggarwal}\affiliation{Panjab University, Chandigarh 160014, India}
\author{Z.~Ahammed}\affiliation{Variable Energy Cyclotron Centre, Kolkata 700064, India}
\author{I.~Alekseev}\affiliation{Alikhanov Institute for Theoretical and Experimental Physics, Moscow, Russia}
\author{J.~Alford}\affiliation{Kent State University, Kent, Ohio 44242, USA}
\author{C.~D.~Anson}\affiliation{Ohio State University, Columbus, Ohio 43210, USA}
\author{A.~Aparin}\affiliation{Joint Institute for Nuclear Research, Dubna, 141 980, Russia}
\author{D.~Arkhipkin}\affiliation{Brookhaven National Laboratory, Upton, New York 11973, USA}
% ...

\collaboration{STAR Collaboration}\noaffiliation
 
\begin{abstract}     
Previous experimental results based on data ($\sim15$ million events) collected by the STAR detector at RHIC
suggest event-by-event charge separation fluctuations perpendicular to
the event plane in non-central heavy-ion collisions. Here we
present the correlator previously used split into its two component
parts to reveal correlations parallel and perpendicular to the
event plane. The results are from a high statistics 200 GeV Au+Au
collisions data set (57 million events) collected by the STAR experiment.
We explicitly count units of charge separation from which we find
clear evidence for more charge separation fluctuations perpendicular
than parallel to the event plane. We also employ a modified correlator
to study the possible ${\cal P}$-even background in same and opposite charge
correlations, and find that the ${\cal P}$-even background may largely
be explained by momentum conservation and collective motion.
\end{abstract} 
  
\maketitle  

\section{Introduction}
Parity violation represents a preference of handedness in nature.  It
may be violated globally or locally.  In the global sense, the weak
interactions of the standard model are parity odd~\cite{LeeYang:1956} while
the strong interactions are parity even at vanishing temperature and isospin density~\cite{WhittenVafa:1984}.  However,
it has been found possible for parity to be violated locally in
microscopic domains in QCD at finite temperature~\cite{Lee:1973,LeeWick:1974}.
Parity-odd (${\cal P}$-odd) domains in QCD are the consequence of topologically non-trivial
configurations of gauge fields.  A particular domain may be
characterized by its topological charge.  States with positive and
negative topological charge both violate parity but with an opposite observable pattern. 
Only states with zero topological charge conserve parity.
The global conservation of parity in QCD occurs since positive and negative topological charge
states are equally probable in nature.

The hot, dense, and deconfined QCD matter produced at RHIC is a natural place to
study such ${\cal P}$-odd domains.  A hypothesis has
been made stating that these ${\cal P}$-odd domains might be
observable in heavy-ion collisions.  The so called \textit{Chiral Magnetic
Effect} (CME) states that ${\cal P}$-odd domains can interact with the very large
magnetic fields in non-central collisions yielding charge-separation
parallel to the system's orbital angular momentum~\cite{Kharzeev:2006,KZ:2007,KMcLW:2008,FKW:2008}.  This can be viewed as
the creation of an electric dipole moment vector perpendicular to the reaction plane 
(the plane which contains the impact parameter and the beam momenta).
In practice, the estimated reaction plane is called the event plane.

For a given sign of topological charge and magnetic field, the sign of the
electric dipole moment produced by the CME is also fixed (parity violation).
However, positive and negative topological charges are equally likely
and cannot be distinguished on an event-by-event basis.  One therefore expects the
CME to instead manifest itself in an experiment as charge-separation \textit{fluctuations} perpendicular to the
reaction plane.

Previous STAR results based on 15 million Au+Au events at 200 GeV from RHIC 2004 data 
reported an experimental observation of the charge-separation fluctuations possibly providing an evidence 
for the CME~\cite{STAR:PRL, STAR:PRC}.
A comparable signal was observed by the ALICE experiment with 13 million Pb+Pb events at 2.76 TeV~\cite{ALICE:PRL}.
Besides higher statistics analyzed, this article complements the previous publications in two principle ways.  First, we present the correlator, $\mean{\cos(\phi_{\alpha}+\phi_{\beta}-2\Psi_{\rm RP})}$, split into its in-plane and out-of-plane components (See Eq.~\ref{eq:ThreePoint}).  Second, we compare the correlator previously used to a modified correlator.  The comparison enables a better
understanding of the suppression of opposite w.r.t.~same charge correlations measured with $\mean{\cos(\phi_{\alpha}+\phi_{\beta}-2\Psi_{\rm RP})}$.

This article is divided into six sections.  In Sec. II we describe the STAR experimental setup and data taking conditions used in
this analysis.  In Sec. III we describe the methodology of the analysis including the definitions of correlations measured.
In Sec. IV we discuss the systematic uncertainties which mainly arise due to event plane resolution uncertainties
in the modified correlator.  In Sec. V we
present our results.  Finally in Sec. VI we summarize our results.

\section{Experimental Setup and Data Taking}
\label{sec:setup}

In this analysis, 57 million minimum bias events taken by the STAR detector~\cite{STAR:2003} at RHIC during
the 2007 Au+Au run at \snn$=200$ GeV are used.  A hadronic minimum bias
trigger was formed by requiring a spectator neutron signal above the threshold value in both zero-degree calorimeters (ZDC).
Two ZDC shower maximum detectors (ZDC-SMD) measure the spectator neutron spatial distributions.  The ZDC-SMDs are
located in the beam rapidity regions~\cite{STAR:ZDC}.  Charged particles were tracked primarily with the
STAR Time Projection Chamber (TPC).  Tracks are retained if their transverse momentum and pseudorapidity are in the range $0.15 < p_T < 2$ GeV/$c$ and $|\eta| < 1.0$, respectively.
Event and track cuts are chosen to be the same as in the previous STAR
publications on this subject~\cite{STAR:PRL, STAR:PRC}.
Centrality in this data set is determined from the global tracking of charged particles satisfying specific track quality cuts in the pseudorapidity region
$|\eta| < 0.5$ and with the distance of closest approach (DCA) to the primary-vertex less than 3 cm~\cite{STAR:DirectedFlow}.

\section{Method of Analysis}
\label{sec:method}

The correlation function used in our previous publication to search for the CME is given by~\cite{Voloshin:2004}:
\begin{eqnarray}
&& \mean{\cos(\phi_{\alpha}+\phi_{\beta}-2\Psi_{\rm RP})} \nonumber \\
&=& \mean{\cos(\Delta\phi_{\alpha})\cos(\Delta\phi_{\beta}) -
\sin(\Delta\phi_{\alpha})\sin(\Delta\phi_{\beta})} \nonumber \\
&=& [\mean{v_{1,\alpha}v_{1,\beta}} + B_{\rm IN}] -[\mean{a_{1,\alpha}a_{1,\beta}} + B_{\rm OUT}]. \label{eq:ThreePoint}
\end{eqnarray}
The averaging is done over all particles in an event and over all events.
$\phi_{\alpha}$ and $\phi_{\beta}$ are the azimuthal angles of particles $\alpha$ and $\beta$, respectively.  $\Psi_{\rm RP}$ represents the azimuthal angle of the reaction plane and $\Delta\phi = (\phi-\Psi_{\rm RP})$.  $B_{\rm IN}$ and $B_{\rm OUT}$ represent 
${\cal P}$-even background
processes which may or may not cancel.  $v_1$ and $a_1$ are the first harmonic coefficients
in the Fourier decomposition of
the azimuthal distribution of particles of a given transverse momentum and rapidity:
\begin{eqnarray}
\frac{dN_{\alpha}}{d\phi} &\propto& 1 + 2v_{1,\alpha}\cos(\Delta\phi) + 2v_{2,\alpha}\cos(2\Delta\phi) + ... \nonumber \\
&+& 2a_{1,\alpha}\sin(\Delta\phi) + 2a_{2,\alpha}\sin(2\Delta\phi) + ...
\label{eq:FourierExp}
\end{eqnarray}
Conventionally we call $v_1$ ``directed flow" and $v_2$ ``elliptic flow".

We refer to Eq.~\ref{eq:ThreePoint} as the \textit{three-point correlator}.  Since the reaction plane is not directly measurable we estimate it
using event planes.  The event planes are calculated from the particle
distributions themselves,
\begin{equation}
\Psi_n = \frac{1}{n} \tan^{-1}\left[ \frac{\sum w_i \sin(n\phi_i)}{\sum w_i \cos(n\phi_i)} \right], \label{eq:PsiEq}
\end{equation}
where $n$ is the harmonic and $w_i$ is a weight for each particle $i$ in the sum~\cite{ArtSergei}.  The weight is chosen to be the $p_T$ of the particle itself when $0.15<p_T<2$ GeV/$c$ to
increase the event plane resolution.
Above 2 GeV/$c$, the weight is set to 2.
A first harmonic event plane ($\Psi_1$) is obtained
from the spectator neutron distributions detected in the STAR ZDC-SMD~\cite{STAR:ZDC}.  This type of
event plane exploits the directed flow of spectator neutrons measured at very forward rapidity.
A second harmonic event plane ($\Psi_2$) is obtained by exploiting the
large elliptic flow of charged hadrons measured at mid-rapidity in the TPC,
and is also called ``the participant plane".
The difference between $\Psi_1$ and $\Psi_2$ mainly lies in
the event-by-event fluctuations~\cite{participant_plane}, and presents
a major systematic uncertainty in this paper.

As the CME causes charge separation
fluctuations perpendicular to the reaction plane, it is the sine
part of the three-point correlator which is sensitive to the CME.  Note that the cosine part
serves to establish a reference or baseline to the measurement since
both parts are equally sensitive to backgrounds unrelated to the reaction plane.  In this article we present
measurements of both parts.

The three-point correlator weights different azimuthal regions of charge
separation differently, i.e.~oppositely charged pairs which are
emitted azimuthally at $90^\circ$ from the event plane (maximally
out-of-plane) are weighted more heavily than those emitted only a few degrees
from the event plane (minimally out-of-plane).  We wish to modify the three-point correlator such that all azimuthal regions of charge separation are weighted identically.
The modification, in particular, allows us to better understand the source of the suppression of opposite charge correlations
seen previously~\cite{STAR:PRL, STAR:PRC}.

This may be done by first rewriting Eq.~\ref{eq:ThreePoint} as
\begin{eqnarray}
&&\mean{\cos(\phi_{\alpha}+\phi_{\beta}-2\Psi_{\rm RP})} = \nonumber \\
&&\mean{(M_{\alpha}M_{\beta}S_{\alpha}S_{\beta})_{\rm IN}} -
\mean{(M_{\alpha}M_{\beta}S_{\alpha}S_{\beta})_{\rm OUT}},
\label{eq:MMSS}
\end{eqnarray}
where $M$ and $S$ stand for the absolute magnitude ($0\leq M \leq 1$) and sign ($\pm 1$) of the sine or cosine function, respectively.  IN represents the cosine part of Eq.~\ref{eq:ThreePoint} (in-plane) and OUT
represents the sine part of Eq.~\ref{eq:ThreePoint} (out-of-plane).

To study the dependence of expression \ref{eq:MMSS} on $M$ we compare the correlations obtained to those of a reduced version:
\be
\left(\frac{\pi}{4}\right)^2\left({\mean{S_{\alpha}S_{\beta}}_{\rm IN}-\mean{S_{\alpha}S_{\beta}}_{\rm OUT}}\right) \equiv {\rm msc}.
\label{eq:msc}
\ee
We refer to Eq.~\ref{eq:msc} as a modulated sign correlation (msc).  The transition from Eq.~\ref{eq:MMSS} to Eq.~\ref{eq:msc} can be seen with the following two reductions:
$\mean{MS} \rightarrow \mean{M}\mean{S}$ and $\mean{M_{\alpha}M_{\beta}} \rightarrow \mean{M}^2$.  $\mean{S_{\alpha}S_{\beta}}$ may be written as sum of terms involving Fourier coefficients of which only the odd harmonics contribute.  The common coefficient for all contributions is $(4/\pi)^2$ with a pre-factor of $1/n$, where $n$ is the order of the harmonic.  For this reason we choose $\mean{M_{\rm IN}}^2=\mean{M_{\rm OUT}}^2=(\pi/4)^2$.  With this choice, the msc is also given by the far right hand side
of Eq.~\ref{eq:ThreePoint} when the $n=1$ Fourier coefficients dominate over the other odd coefficients.
The msc differs from the three-point correlator in the inclusion of higher harmonics 
and in the removal of ''magnitude correlations" (correlations of $M_{\alpha}$
with $M_{\beta}$), which are of ``non-flow" origin.  
By non-flow, we mean the correlations not related to the orientation of the reaction plane.
Since the msc is not a pure harmonic, its event plane resolution correction is also not generally localized within one harmonic.  However, we are justified in using the same correction so long as $a_1 \gg a_n/n$ and $v_1 \gg v_n/n$ ($n=3,5,7...$) or if at least $a_n$ fluctuations are similar in magnitude as $v_n$ fluctuations.  We correct both the three-point correlator as
well as the msc with a second harmonic sub-event plane resolution, $\left< \cos(2(\Psi_{a}-\Psi_{b}))\right>^{1/2}$, where $\Psi_a$ and $\Psi_b$ are the event plane angles in sub-event $a$ and $b$,
respectively.  We discuss the systematic uncertainties associated with this correction applied to the msc in Sec. IV.

For a known reaction plane, $\mean{S_{\alpha}S_{\beta}}$ is given
simply by the net number of particle pairing combinations divided by the total number
of combinations.  The net number of particle pairing combinations is defined as the difference in the
number of same side and opposite side combinations.
For the same charge in-plane correlations we have
\be
\mean{S_{\alpha}S_{\beta}}_{\rm IN} =
\frac{N_{\delta}^{\rm L}\left(N_{\delta}^{\rm L}-1\right) + N_{\delta}^{\rm R}\left(N_{\delta}^{\rm R}-1\right) -
  2N_{\delta}^{\rm L}N_{\delta}^{\rm R}}{N_{\delta}\left(N_{\delta}-1\right)},
\label{eq:NetCombosSS}
\ee
where $\delta=+$ for $\alpha\beta=++$, and $\delta=-$ for $\alpha\beta=--$.
For opposite charge in-plane correlations we have
\be
\mean{S_{\alpha}S_{\beta}}_{\rm IN} =
\frac{N_{+}^{\rm L}N_{-}^{\rm L} + N_{+}^{\rm R}N_{-}^{\rm R} -
  N_{+}^{\rm L}N_{-}^{\rm R} - N_{-}^{\rm L}N_{+}^{\rm R}}{N_{+}N_{-}}.
\label{eq:NetCombosOS}
\ee
Here $N$ stands for the number of particles detected either on the
left ($\rm L$) or right ($\rm R$) of the perpendicular to the
reaction plane in the transverse plane and with a positive (+) or
negative (-) charge.  For out-of-plane
correlations one simply replaces $\rm L$ and $\rm R$ with $\rm T$ and $\rm B$ (top and bottom
of the reaction plane in the transverse plane).

To avoid self correlations where a particle is trivially correlated with an event plane calculated in the same
particle pool, we use two equal multiplicity sub-events to calculate the msc.  The sub-events
are statistically independent with random particle assignments.
With sub-events, the azimuthal locations ($\rm T,\rm B,\rm L,\rm R$) of the particles from
one sub-event are calculated with respect to the sub-event plane from the other sub-event.

\subsection{Charge separation counting}

Units of in-plane and out-of-plane charge separation are defined as
\begin{eqnarray}
\Delta Q_{\rm IN} = \left(N_+^{\rm L} - N_-^{\rm L}\right) -\left(N_+^{\rm R} - N_-^{\rm R}\right),\nonumber\\
\Delta Q_{\rm OUT} = \left(N_+^{\rm T} - N_-^{\rm T}\right) - \left(N_+^{\rm B} -
N_-^{\rm B}\right),
\label{eq:DeltaQ}
\end{eqnarray}
respectively.
They can be understood as the net charge on one side of the event plane minus the net
charge on the opposite side of the event plane. The choice of sign for $\Delta Q$ is
irrelevant here.

Equation \ref{eq:msc} like Eq.~\ref{eq:ThreePoint} is sensitive to ${\cal P}$-even correlations not related
to charge separation.  With the aim of separating out the simplest
effects of charge separation fluctuations from other ${\cal P}$-even
backgrounds we express the msc in terms of states of
observed charge separation $\Delta Q$.  By simplest we mean the contribution to Eq.~\ref{eq:msc} which
arises from different $\Delta Q_{\rm OUT}$ and $\Delta Q_{\rm IN}$ probability distributions.
We rearrange the msc into two terms:
\begin{eqnarray}
  &{\rm msc} = \Delta{\rm msc} + \Delta N & \label{eq:mscSplit}\\
  &\Delta {\rm msc} = &\nonumber \\
 &\frac{1}{N_{\rm E}}\sum_{\Delta Q} \left<N(\Delta
Q)\right>\left[{\rm msc}_{\rm IN}(\Delta Q)-{\rm msc}_{\rm OUT}(\Delta Q)\right]  \label{eq:Dmsc}&\\
  &\Delta N = &\nonumber \\
&\frac{1}{N_{\rm E}}\sum_{\Delta Q} \left<{\rm msc}(\Delta
    Q)\right>\left[N_{\rm IN}(\Delta
  Q)-N_{\rm OUT}(\Delta Q)\right], \label{eq:DN}&
\end{eqnarray}
where the sum goes over all observed units of charge-separation.
$N_{\rm E}$ stands for the total number of events.
$N_{\rm IN}(\Delta Q)$ stands for the number of events with $\Delta Q$ units of in-plane
charge separation, and ${\rm msc_{IN}}(\Delta Q)$ stands for the $\langle \rm{msc} \rangle$ in those events. 
The averages, $\left<N(\Delta Q)\right>=(N_{\rm IN}(\Delta Q)+N_{\rm OUT}(\Delta Q))/2$ 
and $\left<{\rm msc}(\Delta Q)\right>=({\rm msc}_{\rm IN}(\Delta Q)+{\rm msc}_{\rm OUT}(\Delta Q))/2$, 
represent an average over in-plane and out-of-plane parts.

\begin{figure}
  \centering
  \includegraphics[width=.2\textwidth]{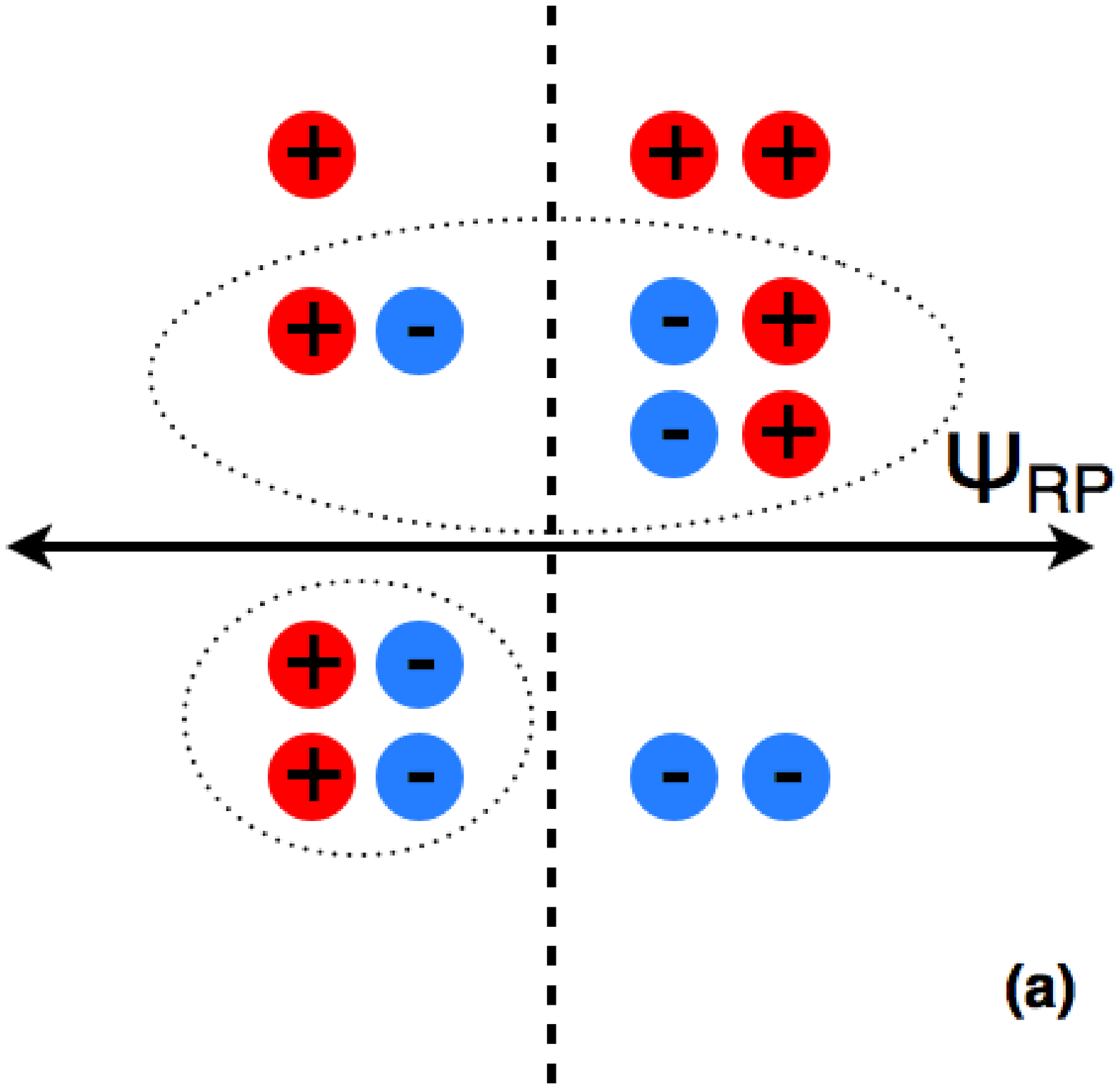}
  \includegraphics[width=.2\textwidth]{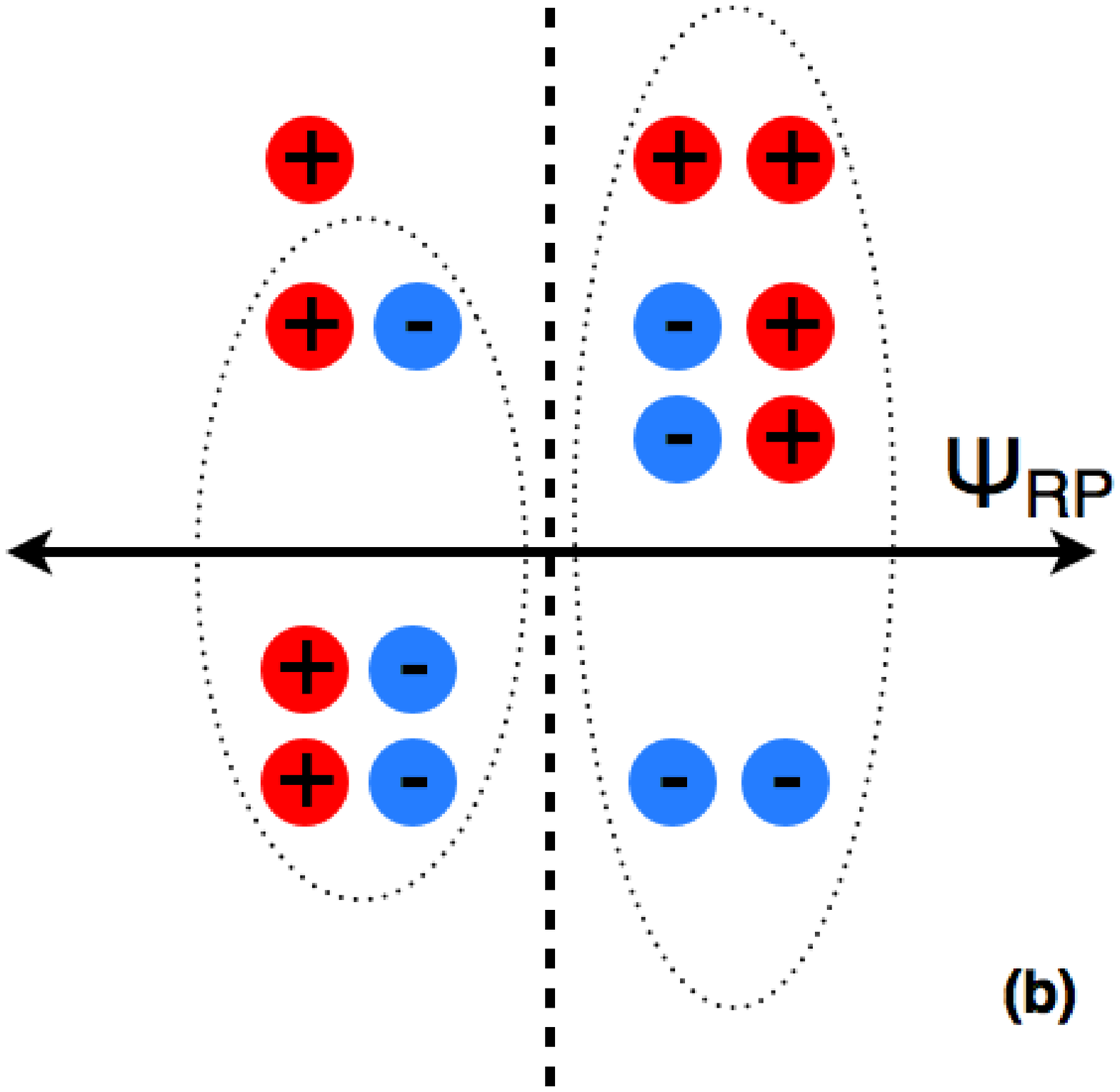}
  \caption{ (Color online)
    Example of a charge configuration with the underlying neutral pairing enclosed by dotted ovals. (a) shows the
    procedure for counting $\Delta Q_{\rm OUT}=+5$. (b) shows the same event but with the procedure for counting $\Delta Q_{\rm IN}=+1$.
}
  \label{fig:DeltaQPic}
\end{figure}

A given
$\Delta Q$ state will be a superposition of many different configurations or sub-states.  The
sub-states may be described in terms of an underlying neutral
pairing of particles plus the residual net charge on each side (T, B, L or R).
The underlying neutral pairs are formed by pairing up
positively and negatively charged particles on a particular side
until only a residual net charge remains.
The residual net charge in a given $\Delta Q$ bin may also be arranged in several ways.
For example, consider the state $\Delta Q_{\rm IN}=+2$.
The residual net charge is +2 units.  One sub-state is the case when the left side has a net charge of +2 and the
right side has a net charge of 0 (neutral).  Another sub-state is formed with a net charge of -2 on the right and 0
on the left.  The other sub-state occurs when the left side has +1 and the right side has  -1 units of net charge.
The idea of charge separation counting is illustrated in Fig.~\ref{fig:DeltaQPic}.
Both the underlying neutral pairing and the residual net charge contribute to the overall configuration within each
$\Delta Q$ bin.

The right hand side of Eq.~\ref{eq:mscSplit} is composed of two terms. The first term, $\Delta {\rm msc}$, is
sensitive to the difference between in-plane and out-of-plane $\Delta Q$ configurations (${\rm msc}_{\rm IN}(\Delta Q)-{\rm msc}_{\rm OUT}(\Delta Q)$).
The second term, $\Delta N$, is sensitive to the
difference between in-plane and out-of-plane $\Delta Q$ probabilities ($N_{\rm IN}(\Delta Q)-N_{\rm OUT}(\Delta Q)$).
The factor, $N_{\rm IN}(\Delta Q)-N_{\rm OUT}(\Delta Q)$, is of course identical for same and opposite charge correlations.  Therefore,
the difference between same and opposite charge $\Delta N$ correlations is determined exclusively by the
prefactor $\left<{\rm msc}(\Delta Q)\right>$.
If the CME does not significantly alter the charge-separation sub-states it will be isolated
in the $\Delta N$ term of Eq.~\ref{eq:mscSplit}.
In general, both terms could
be affected by a ${\cal P}$-even background and neither is to be regarded as an
isolation of a ${\cal P}$-even background.

The effects of ${\cal P}$-even local charge conservation and momentum conservation coupled with non-zero $v_2$ has
been shown to yield a substantial background to the same and opposite charge correlations of 
Eq.~\ref{eq:ThreePoint}~\cite{Pratt2010,PrattSorren:2011}.  The contribution of this effect to
Eq.~\ref{eq:msc} should be reduced due to the treatment of the magnitudes of the cosine and sine functions.


\subsection{Acceptance effects}

Anisotropic or imperfect detector acceptance may also induce false
correlations.  The STAR detector has nearly complete azimuthal
coverage but nevertheless we apply a re-centering correction~\cite{SV:2008} to
all the event plane calculations.  The correction is done in bins of
centrality, location of collision parallel to the beam axis, and STAR run number which represents a period
of time with constant detector calibrations.
Particles $\alpha$ and $\beta$ in the three-point correlator are also
re-centered.  The effect of this procedure was only found to be sizable in
the most central bins where the signal is small.

\section{Systematic Uncertainties}
We estimate the systematic uncertainties on our measurements by comparing results obtained
using TPC and ZDC-SMD event planes.  The difference between these two measurements  
forms our estimate of the systematic uncertainty.  
This estimate is shown in the shaded bands in Figs.~\ref{fig:Three_point} and \ref{fig:msc_split}.  
For the three-point correlator, these values characterize non-flow uncertainties in the reaction plane reconstruction.  
For the $\Delta {N}$ and $\Delta {\rm msc}$ terms the values characterize both non-flow uncertainties in the reaction plane 
reconstruction as well as uncertainties in applying the $2^{nd}$ harmonic event plane resolution to the ${\rm msc}$. 

Other systematic uncertainties were studied extensively in the previous publications on this subject \cite{STAR:PRL, STAR:PRC}.  All were shown to be negligible compared to the uncertainty in determining the reaction plane.
The shaded bands in the figures here represent the same uncertainty determined by a comparison of measurements with $1^{st}$ and $2^{nd}$ harmonic event planes.

For the simplified case of pure elliptic flow ($v_2 > 0$) + an added CME signal ($|a_1| > 0$), we have also verified through Monte Carlo simulations that the msc with respect to the sub-event planes corrected by the sub-event plane resolution is equivalent to the msc with respect to the reaction plane.

\section{Results}
\label{sec:results}

\begin{figure}[ht]
  \includegraphics[width=.44\textwidth]{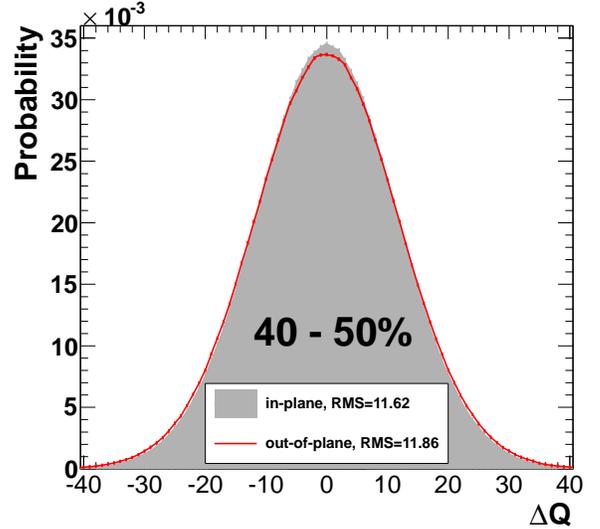}
  \caption{(Color online) Sample $\Delta Q$ distributions for the
    40-50$\%$ centrality Au+Au collisions at \snn$=200$ GeV.  Not corrected for event plane resolution.
    The statistical uncertainties of the RMS values are negligible compared with the difference
$\Delta {\rm RMS}$, as shown in detail in Fig.~\ref{fig:RMS}.}
  \label{fig:DQ}
\end{figure}

\begin{figure}[ht]
  \includegraphics[width=.44\textwidth]{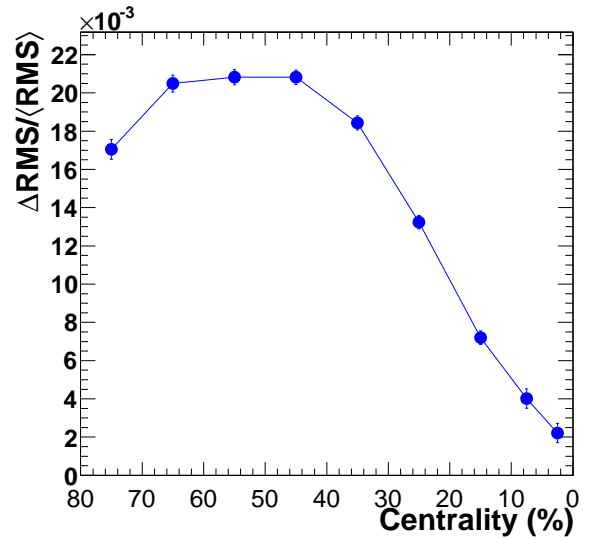}
  \caption{(Color online) $\Delta {\rm RMS^{\Delta Q}}/\mean{\rm RMS^{\Delta Q}}$ versus centrality for Au+Au collisions at \snn$=200$ GeV.  Not corrected for event plane resolution.  Errors are statistical only.}
  \label{fig:RMS}
\end{figure}

Larger charge separation fluctuations perpendicular to rather than parallel to the event plane can be seen by
comparing distributions of $\Delta Q_{\rm OUT}$ to $\Delta Q_{\rm IN}$ as shown in Fig.~\ref{fig:DQ} for the $40-50\%$ centrality bin.
Figure \ref{fig:RMS} shows the difference over the mean of the RMS values ($\frac{{\rm RMS}_{\rm OUT}-{\rm RMS}_{\rm IN}}{({\rm RMS}_{\rm OUT}+{\rm RMS}_{\rm IN})/2}$) versus centrality.
The CME will cause wider out-of-plane distributions, however, ${\cal P}$-even processes may also cause the same feature 
(e.g.~the decays of resonances with sizable $v_2$).
Figures \ref{fig:DQ} and \ref{fig:RMS} are not corrected for the event plane resolution,
however they clearly demonstrate larger charge separation
fluctuations perpendicular rather than parallel to the event plane.
Presumably, the difference between in-plane and out-of-plane distributions should be even larger
if the $\Delta Q$ distributions are measured with the true reaction plane.
In this paper, we would continue with other experimental observables to which
the correction for the event plane resolution is easy to apply.

\begin{figure}[ht]
  \includegraphics[width=.44\textwidth]{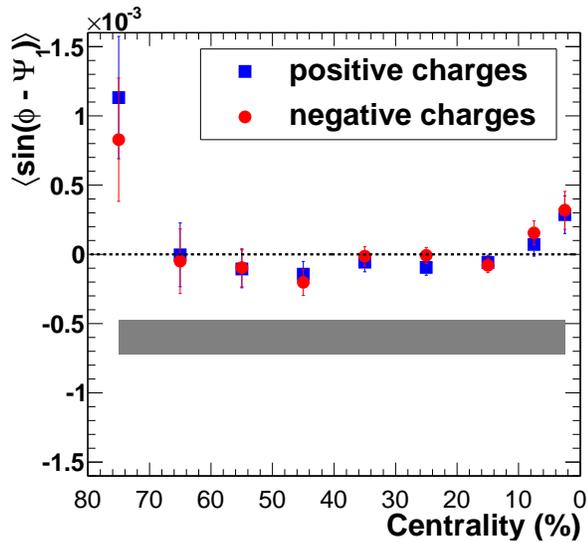}
  \caption{(Color online) $\mean{\sin(\phi_{\alpha}-\Psi_1)}$ for positive and negative charges versus centrality for
Au+Au collisions at \snn$=200$ GeV.  Shaded area represents the systematic
uncertainty for both charge types obtained by comparing correlations from positive and negative pseudorapidity.}
  \label{fig:a1}
\end{figure}

Figure \ref{fig:a1} presents $\mean{\sin(\phi_{\alpha}-\Psi_1)}$ for
positive and negative charges.  Such a measure is sensitive to global parity violation of the strong interactions, i.e. a
preference of charge separation orientation relative to the angular
momentum orientation of
the system.
The results of Fig.~\ref{fig:a1} do not show a significant charge dependence.  The
mean values of both positive and negative charges are less than
$5 \times 10^{-4}$ at the $95\%$ confidence level.  For the most central and peripheral collisions we observe non-zero values for
$a_1$.  However, the values have the same sign for both charge types which is inconsistent with a global violation of parity.

\begin{figure}[ht]
  \includegraphics[width=.44\textwidth]{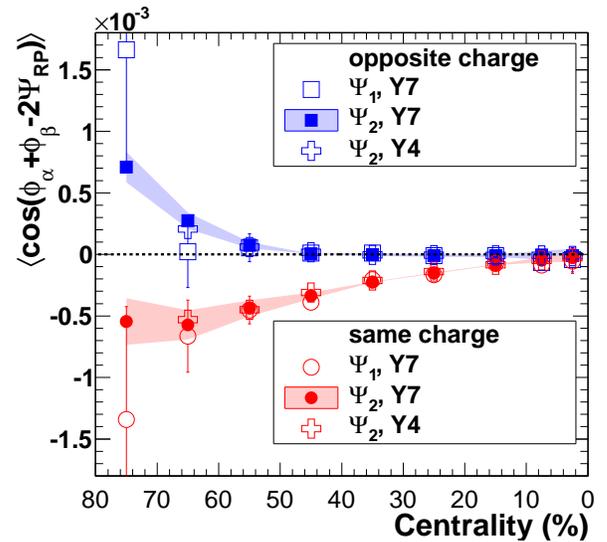}
  \caption{(Color online) Three-point correlator, Eq.~\ref{eq:ThreePoint}, measured with $1^{\rm st}$ and $2^{\rm nd}$ harmonic
    event planes versus centrality for Au+Au collisions at \snn$=200$ GeV.  Shown with crosses are our previous results from the 2004 RHIC run (Y4)
\cite{STAR:PRL, STAR:PRC}.  The Y4 run used
a second harmonic event plane.  Y4 and Y7 $\Psi_2$ results are consistent within statistical errors.
Shaded areas for the $2^{\rm nd}$ harmonic points represent the
systematic uncertainty of the event plane determination.  Systematic uncertainties for the $1^{\rm st}$ harmonic points
are negligible compared to the statistical ones shown.}
  \label{fig:Three_point}
\end{figure}

The three-point correlator measured with $1^{\rm st}$ and $2^{\rm nd}$ harmonic
event planes is shown in Fig.~\ref{fig:Three_point}.
We find consistency between correlations obtained with both
event plane types.  As the pseudorapidity gap between the
ZDC-SMD($\Psi_1$) and the TPC(particles $\alpha$ and $\beta$) is
rather large ($\sim 7$ units in $\eta$) , we find ``direct'' three-particle effects (clusters) to be an
unlikely source for the signal.  This is an indication that the signal
is likely a genuine correlation with respect to the reaction plane.  Also shown for comparison in Fig.~\ref{fig:Three_point} are our previous
results from the 2004 RHIC run \cite{STAR:PRL, STAR:PRC} which are consistent with the current results within statistical errors.

\begin{figure}[ht]
  \includegraphics[width=.44\textwidth]{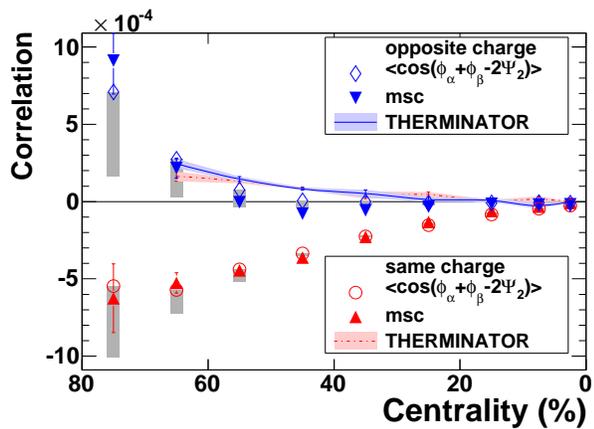}
  \caption{(Color online) Modulated sign correlations (msc) compared to the three-point correlator versus centrality
for Au+Au collisions at \snn$=200$ GeV.  Shown with triangles is the msc, Eq.~\ref{eq:msc}. 
   The systematic uncertainties will be shown in detail in Fig.~\ref{fig:msc_split}.
   Diamonds and circles show the three-point correlator, Eq.~\ref{eq:ThreePoint}, and the grey bars reflect 
   the conditions of $\Delta p_T > 0.15$ GeV/$c$ and $\Delta \eta > 0.15$ applied to the three-point correlator, to be discussed in the text.
   For comparison, the model calculation of THERMINATOR~\cite{Therminator} is also shown.}
  \label{fig:msc_comparison}
\end{figure}

The modulated sign correlations are compared with the
three-point correlator in Fig.~\ref{fig:msc_comparison}.
It is evident that the msc is able to reproduce the same trend as the three-point correlator
although their magnitudes differ slightly.
It is also clear that the correlation magnitude for same charge pairs is larger than for opposite charge pairs for both correlators.
The charge combinations of $++$ and $--$ are consistent with each other for the msc (not shown here),
just like the case for the three-point correlator~\cite{STAR:PRC}.
We also plot the model calculation of THERMINATOR~\cite{Therminator} to be discussed later.

Before any possible interaction with the medium, the CME is expected to generate equal correlation magnitudes for
same and opposite charge pairs.  It was previously supposed that medium suppression of back-to-back phenomena could be responsible for
this \textit{magnitude asymmetry}~\cite{STAR:PRL,STAR:PRC}.  Oppositely charged pairs from the CME may not freeze out
back-to-back, but instead with one of the particles deflected closer to the
event plane due to multiple scattering within the medium.  This is most likely to occur for the particle traversing the largest path length through the medium.
However, when we weight all azimuthal regions of charge separation
equally, as with the msc in Fig. \ref{fig:msc_comparison}, we do not
recover a magnitude symmetry.

\begin{figure}[ht]
  \includegraphics[width=.44\textwidth]{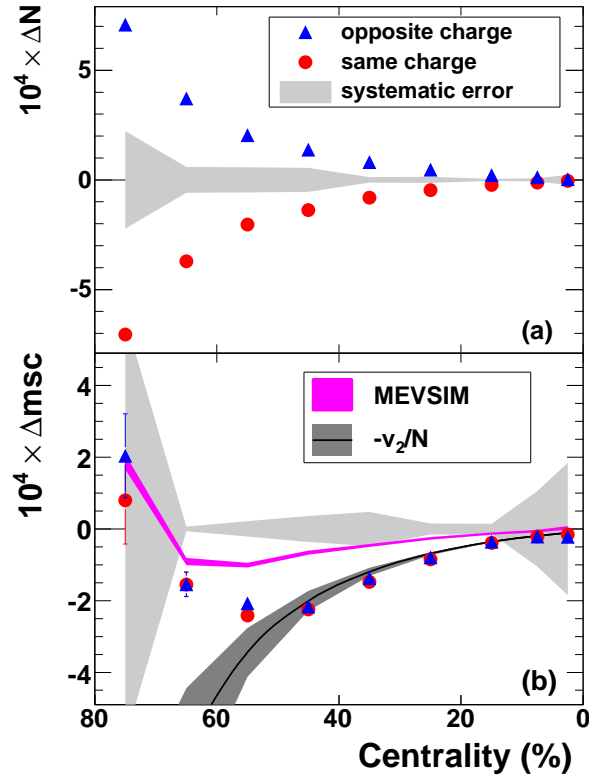}
  \caption{(Color online) The msc split into 2 composite parts versus centrality for Au+Au collisions at \snn$=200$ GeV.  
   Shaded areas represent the systematic uncertainty due to the event plane determination.
   For comparison with the $\Delta \rm{msc}$ term, we also put $-v_{2}/N$ and the model calculation of
   MEVSIM~\cite{MEVSIM}, to be described in the text.   
}
  \label{fig:msc_split}
\end{figure}

The two terms of the msc in Eq.~\ref{eq:mscSplit} are shown in
Fig. \ref{fig:msc_split}.
We observe that same and opposite charge correlations in the $\Delta N$ term have very similar magnitudes, 
but opposite signs for all centrality bins.  This feature is expected from the construction of the
$\Delta N$ term due to the relatively large and
approximately equal positive and negative charge multiplicities.  A model calculation including statistical+dynamical
fluctuations of particle azimuthal distributions should be performed in order to rule out ${\cal P}$-even explanations.
The $\Delta {\rm msc}$ term has a similar magnitude for same and opposite charge correlations,
indicating a charge-independent background for the correlations.
Thus, the source of the magnitude asymmetry between same and opposite charge
correlations about zero as shown in Fig. \ref{fig:msc_comparison} is isolated in the $\Delta {\rm msc}$ term (Note that the sum of
both terms yields the total msc).
To further investigate the source of this background, we plot $-v_2/N$, a simplified estimate of the effect due to 
momentum conservation and elliptic flow~\cite{Bzdak}. 
Here $v_2$ was introduced in Eq.~\ref{eq:FourierExp},
and the values are from Ref.~\cite{FlowLong}. $N$ represents the total number of produced particles,
but in this practice we only counted those within $|\eta|<1$.
$-v_2/N$ well matches the $\Delta \rm{msc}$ term for $0-50\%$ collisions.
MEVSIM is a Monte Carlo event generator, developed for STAR simulations~\cite{MEVSIM}.
A model calculation of MEVSIM with the implementation of $v_2$ and
momentum conservation qualitatively describes the data trend.

\begin{figure}[t]
  \centerline{
    \includegraphics[width=.44\textwidth]{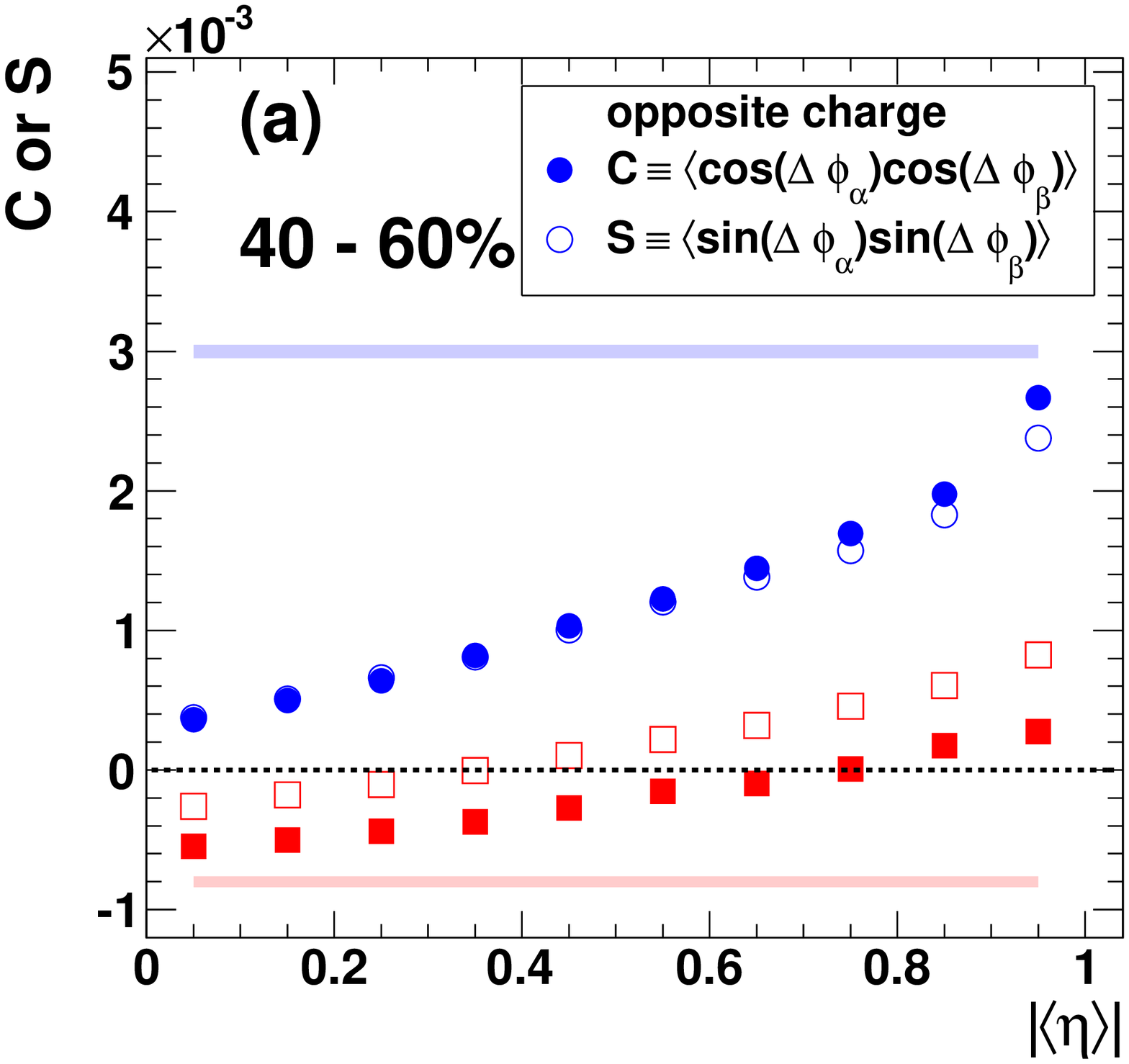}}
  \centerline{
    \includegraphics[width=.44\textwidth]{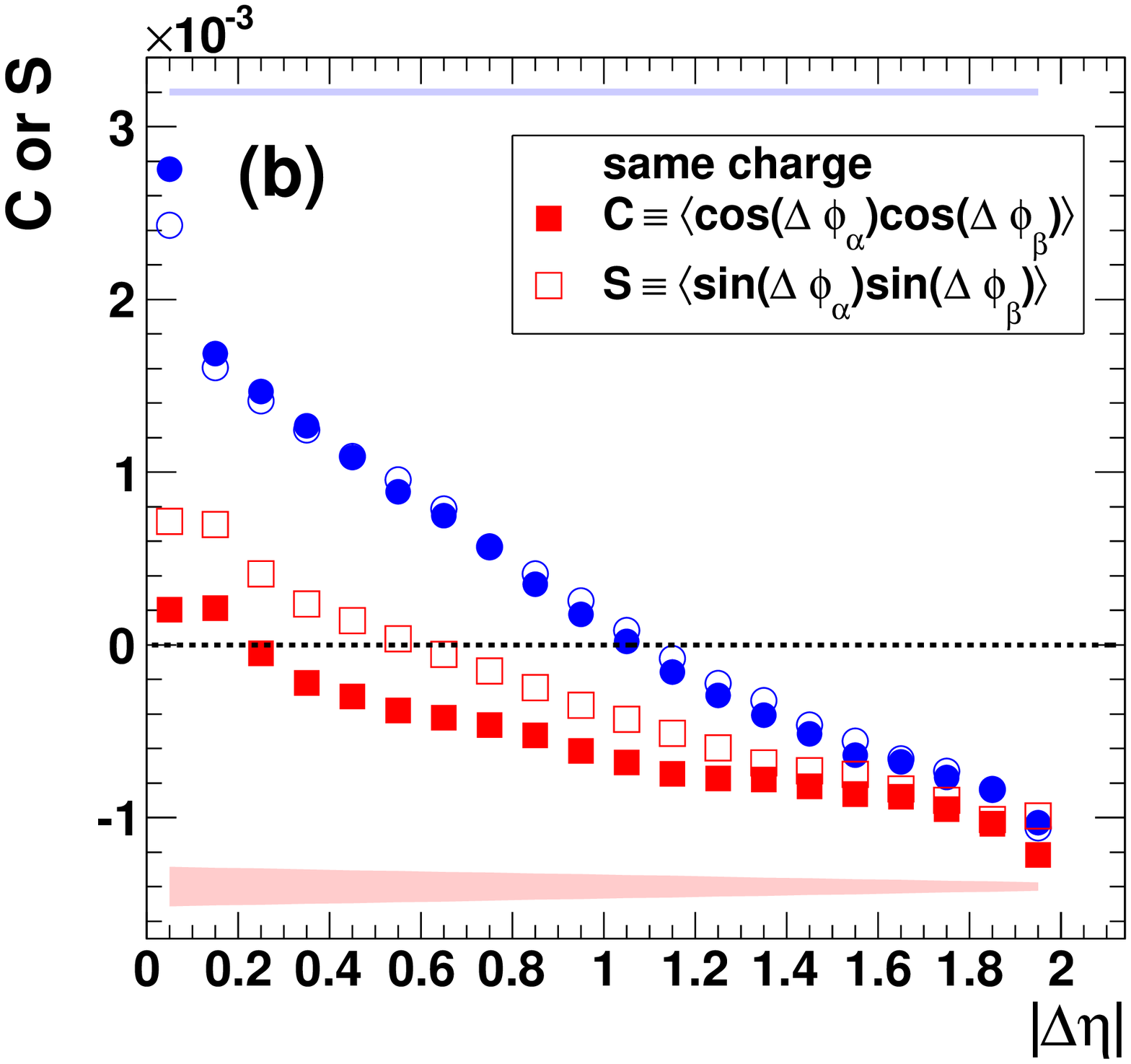}}
  \caption{(Color online) Three-point correlations split up into out-of-plane 
          ($\langle\sin(\Delta \phi_{\alpha})\sin(\Delta \phi_{\beta})\rangle$)
   and in-plane ($\langle\cos(\Delta \phi_{\alpha})\cos(\Delta \phi_{\beta})\rangle$) composite parts for $40 - 60\%$ Au+Au collisions at \snn$=200$ GeV.
   (a) shows the correlations versus $\mean{\eta}=(\eta_{\alpha}+\eta_{\beta})/2$.
   (b) shows the correlations versus $|\Delta \eta|=|\eta_{\alpha}-\eta_{\beta}|$.
   Statistical errors are smaller than the symbol size.
   Systematic errors are given by the shaded bands and apply only to the difference of in-plane and out-of-plane parts.}
  \label{fig:Eta}
\end{figure}

\begin{figure}[t]
  \centerline{
  \includegraphics[width=.44\textwidth]{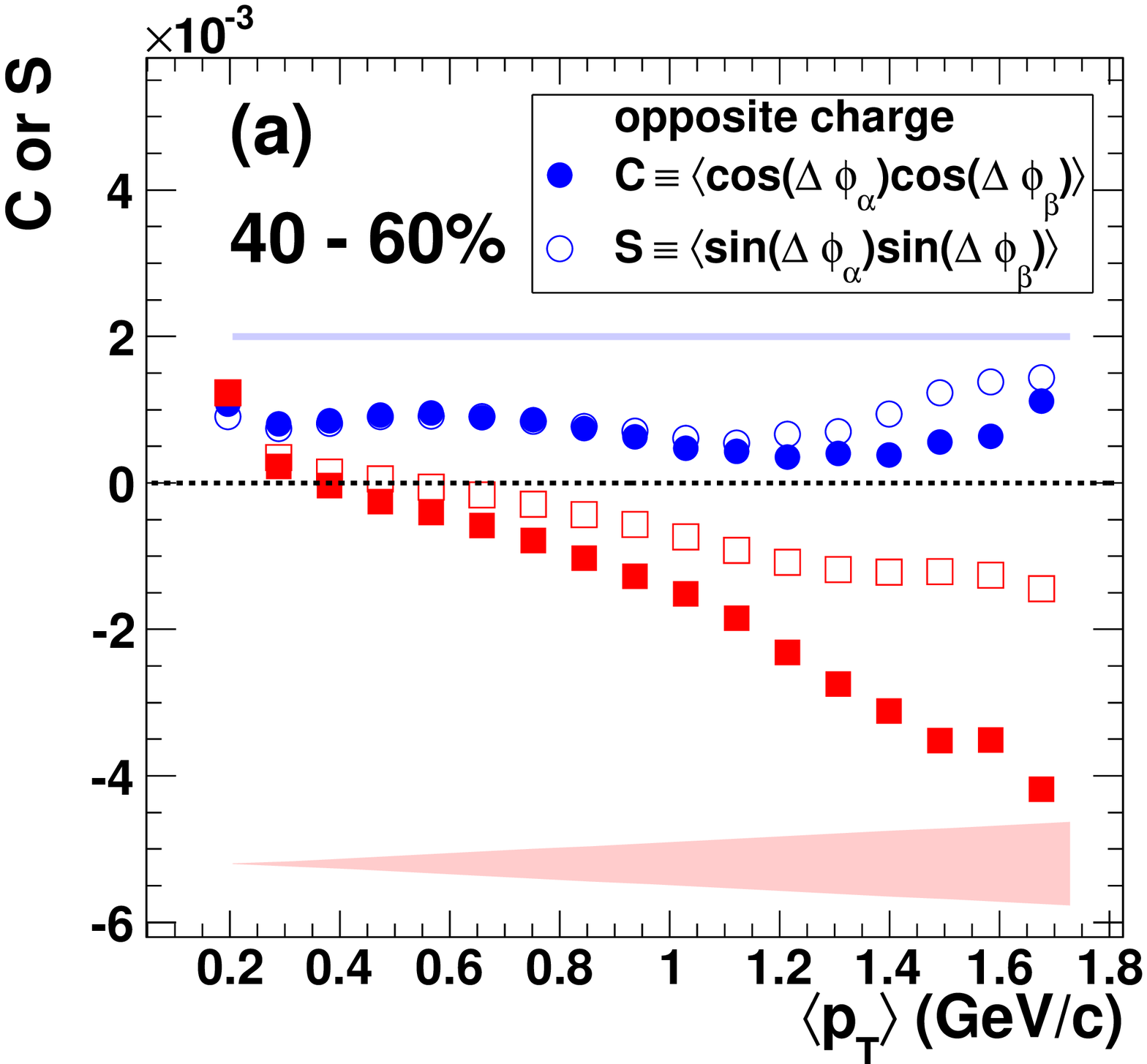}}
  \centerline{
  \includegraphics[width=.44\textwidth]{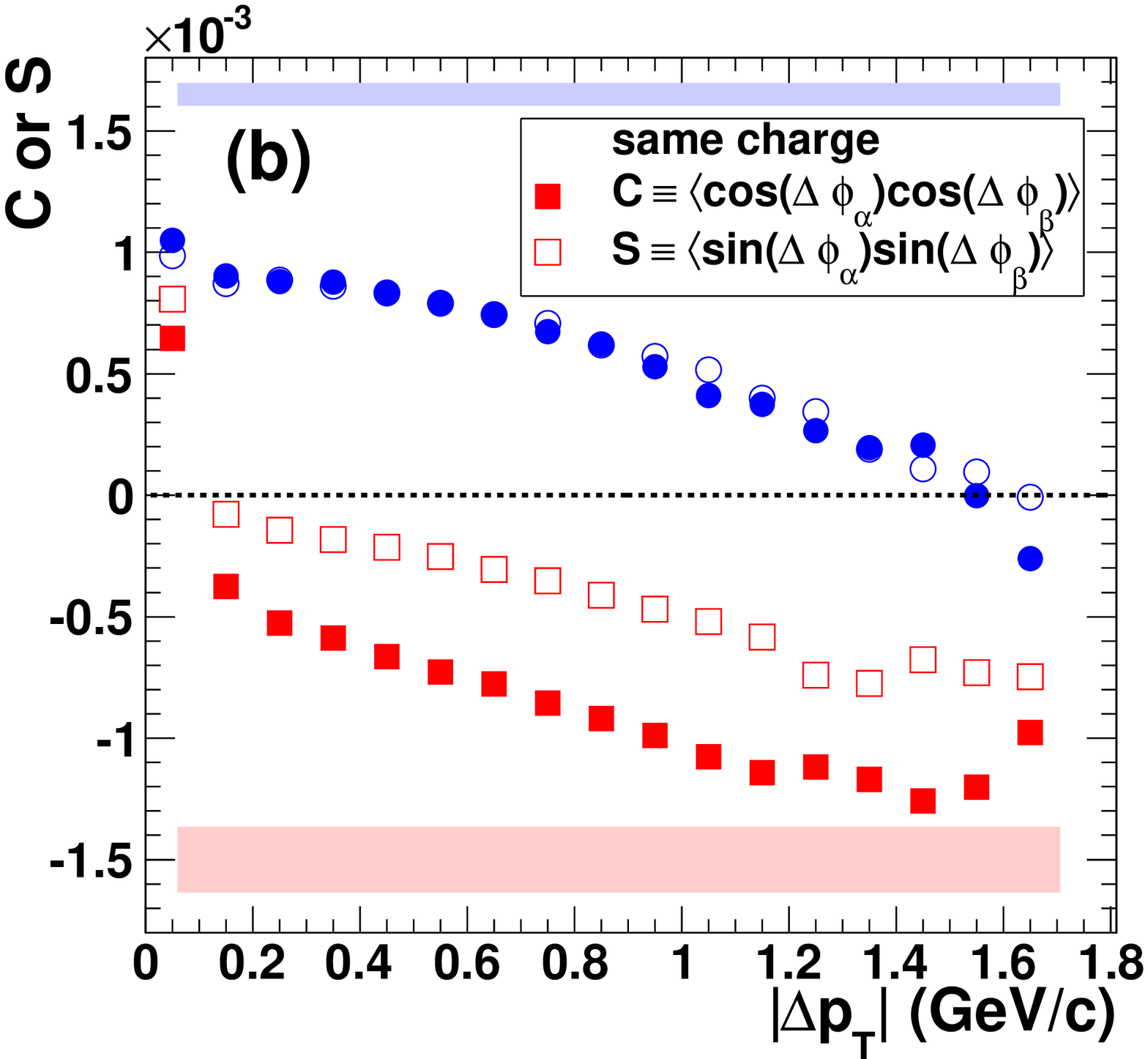}}
  \caption{(Color online) Three-point correlations split up into out-of-plane and
  in-plane composite parts for $40 - 60\%$ Au+Au collisions at \snn$=200$ GeV.
  (a) shows the correlations versus $\mean{p_T}=(p_{T,\alpha}+p_{T,\beta})/2$.
  (b) shows the correlations versus $|\Delta p_T|=|p_{T,\alpha}-p_{T,\beta}|$.
  Statistical errors are smaller than the symbol size.  Systematic errors are given by the
  shaded bands and apply only to the difference of in-plane and out-of-plane parts.}
  \label{fig:Pt}
\end{figure}

We now present the composite parts of the three-point correlation,
Eq.~\ref{eq:ThreePoint}, differentially versus $\eta$ and $p_T$.  Figure \ref{fig:Eta} presents the three-point
correlator versus the average $\eta$ of particles $\alpha$ and $\beta$ ($\mean{\eta}$) and absolute value of the difference
($|\Delta \eta|$).
Figure \ref{fig:Pt} shows the same composite parts versus $\mean{p_T}$ and $\Delta p_T$.
The subtraction of out-of-plane from in-plane composite parts yields the original three-point
correlator while the sum yields a two particle correlation,
$\mean{\cos(\phi_{\alpha}-\phi_{\beta})}$. The split correlations
reveal the underlying ${\cal P}$-even background affecting both composite
parts as each part is sensitive to event plane independent correlations.
We see that in each case the functional shape of in-plane and out-of-plane parts are similar.
The magnitudes of in-plane and out-of-plane parts are more different for same charge pairs.

Femtoscopic correlations at low relative momentum 
which are related to quantum interference (``HBT") and final-state-interactions (Coulomb dominated)
are visible in Figs.~\ref{fig:Eta}a-\ref{fig:Pt}b.
The sharp increase of
the correlation strengths for the lowest bins in Figs.~\ref{fig:Eta}b, \ref{fig:Pt}a, and \ref{fig:Pt}b 
are due to the combination of quantum interference in the same charge channel and the final-state-interactions in both channels.
Low relative momentum in the transverse plane is clearly best visible in Fig.~\ref{fig:Pt}b for low values of $|\Delta p_T|$.  The same phenomena are also
visible for low values of $\mean{p_T}$ since these values best isolate low values of $|\Delta p_T|$.  Low relative momentum along the beam
axis is clearly visible in Fig.~\ref{fig:Eta}b for low values of $|\Delta \eta|$.
The same phenomena are
only visible for the larger values of $\mean{\eta}$ in Fig.~\ref{fig:Eta}a since $\eta$ is signed.  That is, the lowest values of $\mean{\eta}$ contain a substantial fraction of pairs with the opposite sign of $\eta$ and therefore large relative momentum along the beam axis.

We also observe that the positive signal for opposite charge correlations observed in the peripheral bins of 
Fig.~\ref{fig:msc_comparison} ($\mean{\cos(\Delta \phi_{\alpha})\cos(\Delta \phi_{\beta})
-\sin(\Delta \phi_{\alpha})\sin(\Delta \phi_{\beta})}$) is largely found in the kinematic regions of 
Figs.~\ref{fig:Eta}a-\ref{fig:Pt}b where femtoscopic correlations are prominent. 
In Fig.~\ref{fig:msc_comparison}, femtoscopic correlations 
are qualitatively demonstrated by the model calculation of THERMINATOR~\cite{Therminator}.
THERMINATOR is a Monte Carlo event generator designed for studying
of particle production in relativistic heavy-ion collisions, and
includes estimates of the effects of resonance decays, quantum interference, final-state-interactions and collective motions.
To suppress the contribution from femtoscopic correlations, we applied the conditions of $\Delta p_T > 0.15$ GeV/$c$
and $\Delta \eta > 0.15$ to the three-point correlator, shown with the grey bars in Fig.~\ref{fig:msc_comparison}.
Femtoscopic correlations are sensitive to the size of the emission volume at freeze-out \cite{KopylovPod, Goldhaber}.
The difference between in-plane and out-of-plane correlations
in the kinematic region with prominent femtoscopic correlations
can be due to a difference in the emission volumes probed by in and out-of-plane parts.
Such a difference may arrise from an azimuthally anisotropic freeze-out distribution coupled with elliptic flow.

\section{Summary}
Correlations sensitive to charge separation in heavy-ion collisions
have been presented.  Consistency between correlations with
respect to $1^{st}$ and $2^{nd}$ harmonic event planes demonstrates that the signal is likely to be
related to the reaction plane.
Also presented was a reduced version of the three-point
correlation in which all regions of
charge separation are weighted equally.  The same qualitative signal was found to
persist in this scheme as well.  The signal
shown in Fig. \ref{fig:msc_comparison} is
largely determined by the sign ($\pm$) of the cosine and sine
functions in Eq.~\ref{eq:ThreePoint}.

We also explicitly counted units
of charge separation with which we could better understand the source
of the opposite charge suppression.  A parity conserving background,
due to momentum conservation and collective flow, is more likely to explain
the suppression rather than the medium induced back-to-back suppression previously
supposed~\cite{STAR:PRL, STAR:PRC}.
A comparison of the
RMS values for $\Delta Q_{\rm OUT}$ and $\Delta Q_{\rm IN}$ suggests greater
charge separation fluctuations perpendicular to rather than parallel to the event plane.
The CME as well as ${\cal P}$-even processes such as the decays of resonances with sizable $v_2$
may both contribute to this feature.

The differential analysis of the in-plane and out-of-plane parts of the three-point correlator versus $\eta$ and $p_{T}$ reveals
femtoscopic contributions at low relative momentum.  The positive signal in Fig.~\ref{fig:msc_comparison} for opposite charge correlations in peripheral collisions is largely found in the low relative momentum regions of Fig.~\ref{fig:Eta} and Fig.~\ref{fig:Pt}.  This can possibly be explained by the final-state-interactions (${\cal P}$-even)
of different emission volumes probed by in-plane and out-of-plane parts.

Excluding low relative momentum pairs significantly reduces the positive contributions to opposite charge correlations in
Fig.~\ref{fig:msc_comparison}.
However, the difference between same and opposite charge correlations remains largely unchanged and consistent with the
expectations of the CME.  ${\cal P}$-even local charge conservation coupled to elliptic flow modeled by charge balance functions has also been shown
to generate same charge three-point correlations comparable to the observed one~\cite{PrattSorren:2011}.
A careful calculation of the mentioned ${\cal P}$-even backgrounds needs to be made before a further
assessment of the Chiral Magnetic Effect can be made in heavy-ion collisions.

\section*{Acknowledgments}
We thank the RHIC Operations Group and RCF at BNL, the NERSC Center at LBNL and the Open Science Grid consortium for providing resources and support. This work was supported in part by the Offices of NP and HEP within the U.S. DOE Office of Science, the U.S. NSF, the Sloan Foundation, CNRS/IN2P3, FAPESP CNPq of Brazil, Ministry of Ed. and Sci. of the Russian Federation, NNSFC, CAS, MoST, and MoE of China, GA and MSMT of the Czech Republic, FOM and NWO of the Netherlands, DAE, DST, and CSIR of India, Polish Ministry of Sci. and Higher Ed., National Research Foundation (NRF-2012004024), Ministry of Sci., Ed. and Sports of the Rep. of Croatia, and RosAtom of Russia.
  
\end{document}